\documentclass[9pt,twocolumn,twoside]{layout/pnas-new}

\templatetype{layout/pnasresearcharticle} 

\usepackage{braket}
\usepackage{bm}
\usepackage{subdepth}

\begin{document}

\title{Universal Relation between Spectral and Wavefunction Properties at Criticality}

\author[a,b]{Simon Jiricek}
\author[b,c]{Miroslav Hopjan}
\author[d]{Vladimir Kravtsov}
\author[e]{ Boris Altshuler}
\author[a,b]{Lev Vidmar}

\affil[a]{Department of Physics, Faculty of Mathematics and Physics, University of Ljubljana, Jadranska ulica 19, SI-1000 Ljubljana, Slovenia}
\affil[b]{Department of Theoretical Physics, J. Stefan Institute, Jamova cesta 39, SI-1000 Ljubljana, Slovenia}
\affil[c]{Institute of Theoretical Physics, Wroclaw University of Science and Technology, 50-370 Wrocław, Poland}
\affil[d]{International Centre for Theoretical Physics (ICTP), Str. Costiera 11, 34151 Trieste, Italy}
\affil[e]{Physics Department, Columbia University, 538 West 120th Street, New York, NY 10027}

\leadauthor{Jiricek}


\significancestatement{
The important role in physics research is to uncover universal properties of various systems with different microscopic descriptions.
An example of a microscopic model that exhibits paradigmatic properties is a hopping lattice model at weak on-site disorder, known as the Anderson model, which in high dimensions supports diffusive transport, exhibits chaotic quantum dynamics, and has spectral and wavefunction properties governed by random-matrix theory.
Radical counterexamples to this behavior are also known, and one of such cases is the well-known Anderson localization.
Nevertheless, much less is known about the possible universal properties at the boundary between quantum chaos and localization, in which the system exhibits critical behavior.
Here, we conjecture, and confirm using large-scale numerical simulations, a universal relation between the spectral compressibility and the wavefunctions' fractal dimension at criticality.
This result paves way towards searching analogous relations in interacting models, for which such universality has not yet been explored.
}

\authorcontributions{Please provide details of author contributions here.}
\authordeclaration{Please declare any competing interests here.}
\correspondingauthor{\textsuperscript{2}To whom correspondence should be addressed. E-mail: author.two\@email.com}


\begin{abstract}
Quantum-chaotic systems exhibit several universal properties, ranging from level repulsion in the energy spectrum to wavefunction delocalization.
On the other hand, if wavefunctions are localized, the levels exhibit no level repulsion and their statistics is Poisson.  
At the boundary between quantum chaos and localization, however, one observes critical behavior, not complying with any of those characteristics.
An outstanding open question is whether there exist yet another type of universality, which is genuine for the critical point.
Previous work suggested that there may exist a relation between the global characteristics of energy spectrum, such as spectral compressibility $\chi$, and the degree of wavefunction delocalization, expressed via the fractal dimension $D_1$ of the Shannon--von Neumann entropy in a preferred (e.g., real-space) basis.
Here we study physical systems subject to local and non-local hopping, both with and without time-reversal symmetry, with the Anderson models in dimensions three to five being representatives of the first class, and the  banded random matrices as representatives of the second class.
Our thorough numerical analysis supports validity of the simple relation $\chi + D_1 = 1$ in all systems under investigation. Hence we conjecture that it represents a universal property of a broad class of critical models.
Moreover, we test and confirm the accuracy of our surmise for a closed-form expression of the spectral compressibility in the  one-parameter critical manifold of random banded matrices. Based on these findings we derive a universal function $D_{1}(r)$, where $r$ is the averaged level spacing ratio, which is valid for a broad class of critical systems.
\end{abstract}

\dates{This manuscript was compiled on \today}
\doi{\url{www.pnas.org/cgi/doi/10.1073/pnas.XXXXXXXXXX}}

\maketitle
\ifthenelse{\boolean{shortarticle}}{\ifthenelse{\boolean{singlecolumn}}{\abscontentformatted}{\abscontent}}{}

\firstpage[1]{5}
 



Quantum-chaotic systems are generic systems with eigenspectra that can be described by the Wigner-Dyson random matrix theory~\cite{Mehta} and with eigenstates delocalized in essentially any basis. Notable examples of such systems are complex nuclei~\cite{Weidenmuller}, quantum dots~\cite{Beenakker}, quantum billiards~\cite{Nakamura}, weakly disordered electrons in high-dimensional lattices~\cite{Altshuler_Shklovskii}, and interacting many-body systems~\cite{dalessio_kafri_16}.
Oppositely, systems with localization exhibit Poisson eigenvalue statistics, which are drastically different from Wigner-Dyson level statistics, since the former correspond to the absence of level repulsion~\cite{Mehta}.
Eigenstate statistics in such systems are basis-dependent, with localization occurring in some preferred basis.
In case of electrons propagating in lattices with large on-site disorder, the preferred basis is the site-occupation basis.

Here, we focus on the critical boundary between these two classes of systems.
Critical systems are associated with intermediate eigenspectra statistics, i.e., neither of Wigner-Dyson nor Poisson type, and with multifractal eigenstates, i.e., neither delocalized over the  volume nor localized, suggesting that the change in eigenspectra is connected to the change in the extension of eigenstates.

Early studies of the critical eigenspectra were mostly focused on the level statistics at the critical point of the localization transition in the three-dimensional (3$d$) Anderson model, specifically, on the level number variance in a certain spectral window \cite{Altshuler_Shklovskii} and the level spacing distribution function \cite{Shklovskii}. It was shown in these works that the critical level statistics have a hybrid character. They share with Poisson statistics the linear dependence of the level number variance as a function of the mean number of levels in the window, and the simple exponential decay of the level spacing distribution at large gaps between levels. However, at small gaps there is level repulsion in the critical statistics, similar to the Wigner-Dyson case.

The next important step concerned the critical eigenfunction statistics, inspired by the discovery of multifractality by F. Wegner~\cite{Wegner1980inverse}.
It was shown that the critical eigenfunctions are characterized by a set of fractal dimensions $0<D_{q}<1$, which are intermediate between the case of fully extended (ergodic) wavefunctions, where all $D_{q}=1$, and the case of localized wavefunctions, where all $D_{q}=0$ \cite{Mirlin_Evers_Review}.

That said, not all of $D_{q}$ are equally important for the description of critical eigenfunctions.
For physical applications related to pair interaction, e.g., for multifractal superconductivity near the localization transition~\cite{MFSC}, the most relevant is the fractal dimension $D_{2}$.
However, the principle information about the fractal character of a wavefunction is encoded in the fractal dimension $D_{1}$. Considering $N$ as the total volume of the system (e.g., $N$ is the number of lattice sites), $D_1$ determines the volume, $N^{D_{1}}$, of its fractal {\it support set}, defined as the  manifold in space that makes the main contribution to eigenfunction normalization \cite{SS}.

A non-trivial question emerges about a quantitative relationship between the spectral characteristics of critical states and their eigenfunction statistics.
On first glance this question is not legitimate, since the spectrum is basis-invariant while the eigenfunctions are not.
Yet, our experience shows that this relation may exist for physically motivated models.
An evidence is the observation, which became a common wisdom, that Poissonian spectral statistics correspond to localized wavefunctions and Wigner-Dyson statistics correspond to fully ergodic states.
Hence, there must exist a basis (or a manifold of bases), denoted as the preferred basis, in which localization effects are strongest, e.g., the support set in such a basis should have a minimal fractal dimension $D_{1}$. 
In most studies of Anderson localization, the preferred basis coincides with the "natural" one in which the model is formulated, e.g., by locality of disorder potential in this basis.

We note that the preferred basis should be fixed, i.e., independent of a specific realization in a given ensemble of random Hamiltonians.
Thus the basis of eigenstates, in which all states are trivially localized, should be excluded, as it depends on the particular realization of disorder. The states are called "localized" when $D_{1}=0$ in the preferred basis.  

\section*{Conjecture about the universal relation at criticality} \label{sec:conjecture}
The above discussion suggests that, when compared to the spectral properties, the wavefunction properties should be expressed in the preferred basis.
The goal of this paper is to establish a link between the wavefunction and spectral properties at the localization transition in physical models.
Finding their universal relations, whenever they exist, is important since they may establish a hallmark of critical behavior in large but finite systems.

The first observation of a possible relation between the spectral and eigenfunction properties was suggested in 1996 by Chalker, Kravtsov and Lerner~\cite{Chalker_Kr_Ler}, which appeared as an important by-product of a non-trivial theory of "Pechucas gas of levels" developed in~\cite{Chalker_Ler_Sm}.    
It gave rise to the relationship $\chi=(1-D_{2})/2$~\cite{Chalker_Kr_Ler}, which linked the fractal dimension $D_{2}$ to the coefficient of proportionality, the spectral compressibility $\chi$, between the level number variance and the mean number of levels in a spectral window containing critical states.
This result appeared to be valid only for weak multifractality when $(1-D_{q})/q$ is the same for all $q$.
A question, which remained open at that time, was which of the fractal dimensions $D_q$ remains relevant for this type of relationship at an arbitrary strength of multifractality.

The answer  to this question was conjectured by Bogomolny and Giraud~\cite{Bogomol}, who demonstrated that a number of unconventional random matrix models with {\it long-range} hopping obey the modified relationship by Chalker, Kravtsov and Lerner~\cite{Chalker_Kr_Ler}, in which $(1-D_{2})/2$ is replaced by $1-D_{1}$. 
Shortly after, the conjecture was tested on a 2$d$ generalization of a random matrix model with long-range hopping~\cite{Ossipov12}.
Later, it was shown~\cite{Bogo2020, Bogo_Giraud2021} that an important class of Toeplitz and Hankel random matrices exhibit critical level and eigenfunction statistics that approximately obey
the same relation between the spectral compressibility $\chi$ and the fractal dimension $D_{1}$.

A key missing point of the above relations is to establish their relevance in physical models, for which the dominant processes are of short-range nature. Recently, advances in numerical approaches allowed for obtaining a fresh perspective on the critical properties of short-range disordered models, such as the Anderson models in two and three dimensions~\cite{suntajs_prosen_21, suntajs_prosen_23}.
In particular, by comparing the scale-invariant critical dynamics at mid and late times, a recent study observed intriguing relations between the wavefunction and spectral properties at criticality~\cite{Hopjan23b}, resembling the conjectured relations discussed above.

Here, we bridge the gap between exact relations in analytically tractable models and exact numerical results in finite physical models, and we establish validity of a simple relation,
\begin{equation} \label{eq:relation}
    \chi + D_1 = 1\;,
\end{equation}
between the spectral compressibility $\chi$ and the wavefunction fractal dimension $D_1$.
We conjecture that \eqref{eq:relation}, first introduced in Ref.~\cite{Bogomol}, is a universal property of a broad class of critical systems at the boundary between quantum chaos and localization.
By universal property we have in mind the independence of the model's dimensionality and symmetry class, and the locality of physical interactions; see also the discussion below.

We present a thorough numerical analysis to test~\eqref{eq:relation} in Anderson models and the critical power-law random banded models~\cite{PLBRM}. The latter model, though of long-range nature, was the simplest one in which the relation, Eq.~(\ref{eq:relation}), was analytically demonstrated \cite{Bogomol} in the limiting cases. 
It is also the model, in which the local spectrum is singular-continuous, with the local density of states being a random Cantor set \cite{Alt_Kr_Cantor}.
For the Anderson models we focus on three, four and five lattice dimensions with and without time-reversal symmetry. Remarkably, we find in all cases that, to high precision,~\eqref{eq:relation} is fulfilled.
In addition, we confirm that~\eqref{eq:relation} also holds for the critical orthogonal and unitary power-law random banded model, in the entire parameter range.

We then make a step forward and explore validity of the available analytical predictions for $\chi$.
Surprisingly, we observe that $\chi$ in the power-law random banded model of unitary symmetry class can be well described by a simple surmise based on the modified Wigner-Dyson kernel, which exploits the analogy of level statistics with the position statistics of 1$d$ fermions at a finite temperature~\cite{MNS, KrMutta, KrTsvel}.
Exact numerical results establish relevance of  our expression.
They show that the quantitative agreement is of similar accuracy as in the case of the well-known nearest level spacing (gap) distribution, where the Wigner surmise~\cite{Mehta} describes the exact numerical distribution in ergodic systems to high accuracy. We apply the same approach to the averaged gap ratio $r$~\cite{Huse_Oganes} and obtain good agreement with the exact numerical result. 

Finally, we derive a relation between the fractal dimension $D_{1}$ and the average gap ratio $r$, which is valid for a broad class of critical systems.    
We check this relationship for the Anderson models in three, four and five dimensions, as well as for the systems with semi-Poisson statistics~\cite{Bogo2020, Bogo_Giraud2021}. In all the cases we find good agreement of our theory with the corresponding numerical result.

\section*{Models and quantities under investigation}

Below we provide information about the models under investigation.
All of them exhibit a well-established {\it single-particle} eigenstate localization transition, enabling a detailed exploration of their critical properties.
Further details can be found in \textit{Materials and Methods}.

\subsection*{Anderson Models}
\label{subsec:anderson_models}

We start by considering Anderson Models in $d=3,4,5$ spatial dimensions, corresponding to single-particle Hilbert space dimensions $N=L^d$, described by a Hamiltonian of the general form:
\begin{align}
\label{eq:Anderson_ham_general}
    \hat{H} = - \sum_{\braket{\bm{r}_j,\bm{r}_l}} t_{\bm{r}_j,\bm{r}_l} (\hat{c}^{\dagger}_{\bm{r}_j}\hat{c}_{\bm{r}_l} + \mathrm{h.c.}) + \sum_{\bm{r}_j} h_{\bm{r}_j} \hat{n}_{\bm{r}_j},
\end{align} 
where $\hat{n}_{\bm{r}_j}\equiv\hat{c}^{\dagger}_{\bm{r}_j}\hat{c}_{\bm{r}_j}$, $\braket{\bm{r}_j,\bm{r}_l}$ denotes neighboring lattice sites, and $t_{\bm{r}_j,\bm{r}_l}$ denotes the hopping element between those sites.
We choose periodic boundary conditions.
The onsite potentials $h_{\bm{r}_j}$ are, for a given disorder strength $W$, drawn randomly from the uniform ("box") probability distribution, $h_{\bm{r}_j} \in [-W/2,W/2]$.
This class of models is known to undergo a transition from single-particle quantum-chaos to localization for $d>2$ \cite{Anderson_58, Abrahams_79, Tarquini17}. 

In the simplest case of a real Hamiltonian and isotropic lattice, $t_{\bm{r}_j,\bm{r}_l} \equiv t$, with real parameter $t$, time-reversal symmetry is obeyed and the model is expected to agree with predictions of the Gaussian orthogonal ensemble (GOE) for disorder strengths well below the transition. Thus we refer to these models as "orthogonal" Anderson models. The transition points are known~\cite{Slevin} with high accuracy to lie at $W_{c}=16.54 \,\, (3d),$ $W_{c}=34.6 \,\, (4d)$ and $W_{c}=57.3 \,\, (5d)$.

To extend to systems of unitary universality class, we further study Anderson models of broken time-reversal symmetry to which we refer as "unitary" Anderson models, since their statistics is expected to agree with predictions of the Gaussian unitary ensemble (GUE). 
To break time-reversal symmetry, we follow two different approaches. 
In the first approach we consider random hopping phases~\cite{Kawarabayashi,Slevin_Ohtsuki_2016}, i.e.,
\begin{align}
\label{eq:def_randphase}
        t_{\bm{r}_j,\bm{r}_l} \equiv t\, \text{e}^{i \varphi_{jl}}\;,\;\;\;
        \varphi_{jl} \in [0,2\pi]\;,\;\;\;
        \varphi_{lj} = - \varphi_{jl} \;,
\end{align}
for the connected lattice sites $\bm{r}_j$ and $\bm{r}_l$. We study this version of the unitary Anderson model in up to five dimensions. While the transition points for 3$d$ and 4$d$ are known from previous studies to emerge at $W_c=18.83$ and $W_c=37.5$, respectively~\cite{Kawarabayashi,Slevin_Ohtsuki_2016}, we pinpoint the transition in five dimensions at $W_c=61.3$.
For the second approach we dress the hopping term with a complex phase in $d-2$ dimensions to mimic the effect of a magnetic field, see \textit{Materials and Methods} for details.

\subsection*{Power-law random banded matrices}

Further we study power-law random banded (PLRB)  matrices \cite{PLBRM} given by the Hamiltonian
\begin{align}
\label{eq:def_plrb}
    \hat{H}_\text{PLRB} &= \sum_{i,j=1}^N h_{ij} \hat c_i^\dagger \hat c_j,\\ 
    h_{ij} &= h_{ji} = \frac{\mu_{ij}}{[1+\frac{N}{\pi}(\sin(|i-j|\frac{\pi}{N})/b)^{2a}]^{1/2}} \;,
\end{align}
where the matrix $\bm{\mu}$, with matrix elements $\mu_{ij}$, is drawn either from the GOE, yielding the PLRB models of orthogonal universality class (GOE-PLRB), or from the GUE (GUE-PLRB). The sine term ensures periodic boundary conditions.
This model exhibits a localization transition at the critical regime of $a = 1$, which will be the focus of this study. The critical eigenstates exhibit strong multifractality for $b \ll 1$ and weak multifractality for $b \gg 1$ \cite{Mirlin_Evers_Review}.

To ensure criticality of states, we extract eigenvalues and eigenstates in all models from a narrow energy window in the middle of the spectrum, at $W=W_{\rm c}$ in the Anderson models and at $a=1$ in the PLRB models.
In all models we denote the eigenvalue problem as $\hat{H} \ket{E_\alpha}=\varepsilon_\alpha \ket{E_\alpha}$. Following the unfolding procedure described in \textit{Materials and Methods}, we set the mean level spacing $\delta$ to unity to obtain the unfolded eigenvalues $E_\alpha$.

\subsection*{Level number variance and spectral compressibility} 
The level number variance $\Sigma^2$ is a characteristic spectral property defined as 
\begin{align}
    \label{eq:def_level_var}
    \Sigma^2(\Delta) = \braket{ \,n^2\,}_H - \braket{\,n\,}_H^2  \;,
\end{align}
where $n$ denotes the number of levels (i.e.~eigenvalues) lying in an energy interval of width $\Delta$, measured in units of mean level spacing $\delta$, while $\braket{\dots}_H$ denotes the averaging over different Hamiltonian realizations. The spectral compressibility $\chi$ is commonly defined via the relation $\Sigma^2 = \chi \braket{n}_H $. Correspondingly, we define an energy-dependent spectral compressibility $\chi(\Delta)$ for random Hamiltonians as
\begin{align}
    \label{eq:def_rel_level_var}
    \chi(\Delta) = \left\langle \frac{\braket{ \,n^2\,}_H - \braket{\,n\,}_H^2}{\braket{\,n\,}_H} \right\rangle_{\hspace*{-1ex}I}\;,
\end{align}
where $\braket{\dots}_I$ denotes the averaging over different interval positions around the middle of the spectrum, see \textit{Materials and Methods} for further details on the numerical implementation.

\begin{figure}
\centering
\includegraphics[width=\columnwidth]{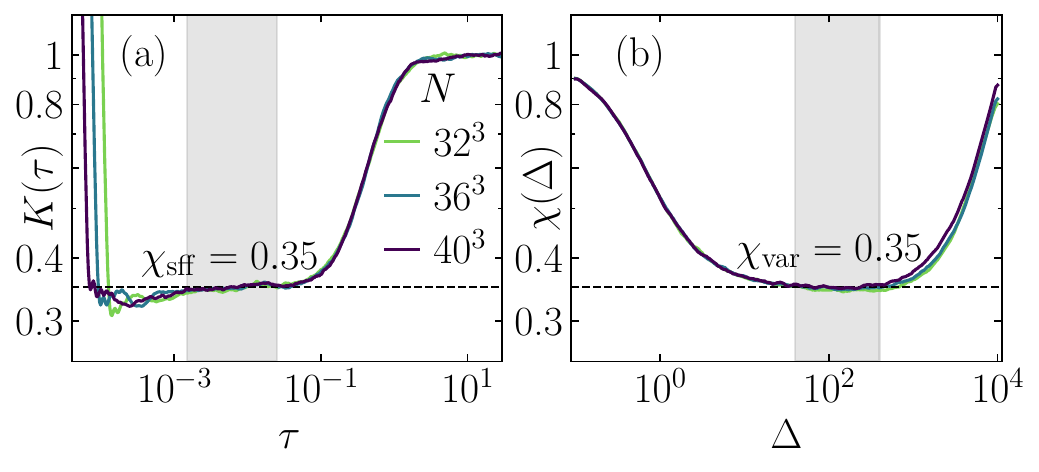}
\caption{Comparison of the two numerical methods to extract the spectral compressibility. 
(a) SFF from~\eqref{eq:def_SFF}, and (b) the energy-depended spectral compressibility from~\eqref{eq:def_rel_level_var} at the critical point $W_c=16.54$ of the 3$d$ orthogonal Anderson model for three different system sizes $N$. 
The plateau values $\chi_\text{sff}$ and $\chi_\text{var}$ in panels (a) and (b), respectively, correspond to the spectral compressibility, extracted in the shaded intervals, .}
\label{fig:fig1}
\end{figure}

The peculiarity of the critical level statistics, first observed in Refs.~\cite{Altshuler_Shklovskii, Kotochigova}, is that in the limit when $N\rightarrow \infty$ is taken first and then $\Delta\rightarrow \infty$, the spectral compressibility $\chi(\Delta)$ tends to a constant value $\chi$,
\begin{align}
    \chi = \lim_{\Delta\,\xrightarrow{}\,\infty} \left(\,\lim_{N\,\xrightarrow{}\,\infty} \chi(\Delta)\right).
\end{align}
For a finite $N$ in the actual numerical calculations this implies a plateau in $\chi(\Delta)$  for $1\ll \Delta\ll N$.
The value of $\chi(\Delta)$ at the plateau, for a sufficiently large $N$, then gives a good estimate of $\chi$, see Fig.~\ref{fig:fig1}(b) for an example.

\subsection*{Spectral form factor and spectral compressibility}

Another commonly investigated spectral measure is given by the spectral form factor (SFF), defined as
\begin{align}
\label{eq:def_SFF}
    K(\tau) = \frac{1}{Z}\left\langle\left|\sum_{\alpha=1}^N \eta(E_\alpha)\text{e}^{-2\pi i E_\alpha \tau}\right|^2\right\rangle_{\hspace*{-1ex}H} \;,
\end{align}
where time $\tau$ is taken in units of the inverse mean level spacing, $Z$ normalizes the long-time value to one and $\eta$ defines a Gaussian filtering function to reduce contributions from the spectral edges~\cite{Suntajs_2020_2}. The definition of the filtering function and further numerical details are elaborated in \textit{Materials and Methods}.

It is known analytically~\cite{Chalker_Kr_Ler} that in the thermodynamic limit, the spectral compressibility $\chi$ is connected to the SFF $K(\tau)$ via 
\begin{align}\label{chi-SFF}
    \chi = \lim_{\tau\,\xrightarrow{}\,0} \left(\,\lim_{N\,\xrightarrow{}\,\infty} K(\tau)\right)\,,
\end{align}
see also \textit{Materials and Methods} for details.
In this work we demonstrate that this relation allows to extract the spectral compressibility at the critical point also from finite-size numerical analyses. This is possible due to the emergence of a scale-invariant plateau in the SFF at intermediate times $\eta_{0}/N\ll \tau\ll 1$ ($\eta_{0}$ is the width of the filtering function) at criticality \cite{suntajs_prosen_21,suntajs_vidmar_22,suntajs_prosen_23,hopjan2023,Hopjan23b}. Since the onset of this plateau drifts to $\tau \xrightarrow{} 0$ as $N \xrightarrow{} \infty $,   the plateau value allows to access the spectral compressibility very accurately even for finite system sizes. An illustrative comparison of the two methods to extract the spectral compressibility is shown in Fig. \ref{fig:fig1} at the critical point of the 3$d$ orthogonal Anderson model.

\subsection*{Fractal dimensions}

We extract the fractal dimensions $D_{q}$ corresponding to the basis $\{\ket{i}\}$ spanned by site-local single-particle states $\ket{i}\equiv \hat c_i^\dagger \ket{\emptyset}$, where $\ket{\emptyset}$ denotes the vacuum state. To this end we first introduce the Shannon--von Neumann entropy~\cite{Shannon} for a given eigenstate $|E_\alpha\rangle$, 
\begin{align}\label{Shannon}
    S_\text{SvN} = - \left\langle\left\langle\sum_{i=1}^N | \braket{i|E_\alpha}|^2 \ln |\braket{i|E_\alpha}|^2 \right\rangle_{\hspace*{-1.2ex}\alpha} \, \right\rangle_{\hspace*{-1.2ex}H}\,,
\end{align}
and the corresponding typical  R\`{e}nyi  entropy~\cite{Renyi},
\begin{align}\label{Renyi}
    S_\text{R}^{(q)}=\frac{1}{1-q}\left\langle\langle\ln P_{q,\alpha}^{-1}\rangle_{\hspace*{-0.2ex}\alpha}\right\rangle_{\hspace*{-0.4ex}H},
\end{align}
where in Eq.~(\ref{Renyi}) the moments of the inverse participation ratios, $P_{q,\alpha}^{-1}$ for $\ket{E_\alpha}$ are defined as 
$P_{q,\alpha}^{-1} = \sum_{i=1}^N |\braket{i|E_\alpha}|^{2q}.$
The averaging over different Hamiltonian realizations, and different mid-spectrum eigenstates within a single realization, is denoted by $\langle... \rangle_{H}$ and $\langle... \rangle_{\alpha}$, respectively.
We then extract the fractal dimension $D_q$ from a fit to the scaling ansatz, 
\begin{align}
\label{eq:scaling_ansatz}
    S_\text{R}^{(q)} = D_{q} \ln N + c_q \;,
\end{align}
in which for the special case $q=1$ the R\`{e}nyi entropy should be replaced by the Shannon--von Neumann entropy.

\section*{Testing the conjecture}

\subsection*{Numerical results}

\begin{figure}
\centering
\includegraphics[width=\columnwidth]{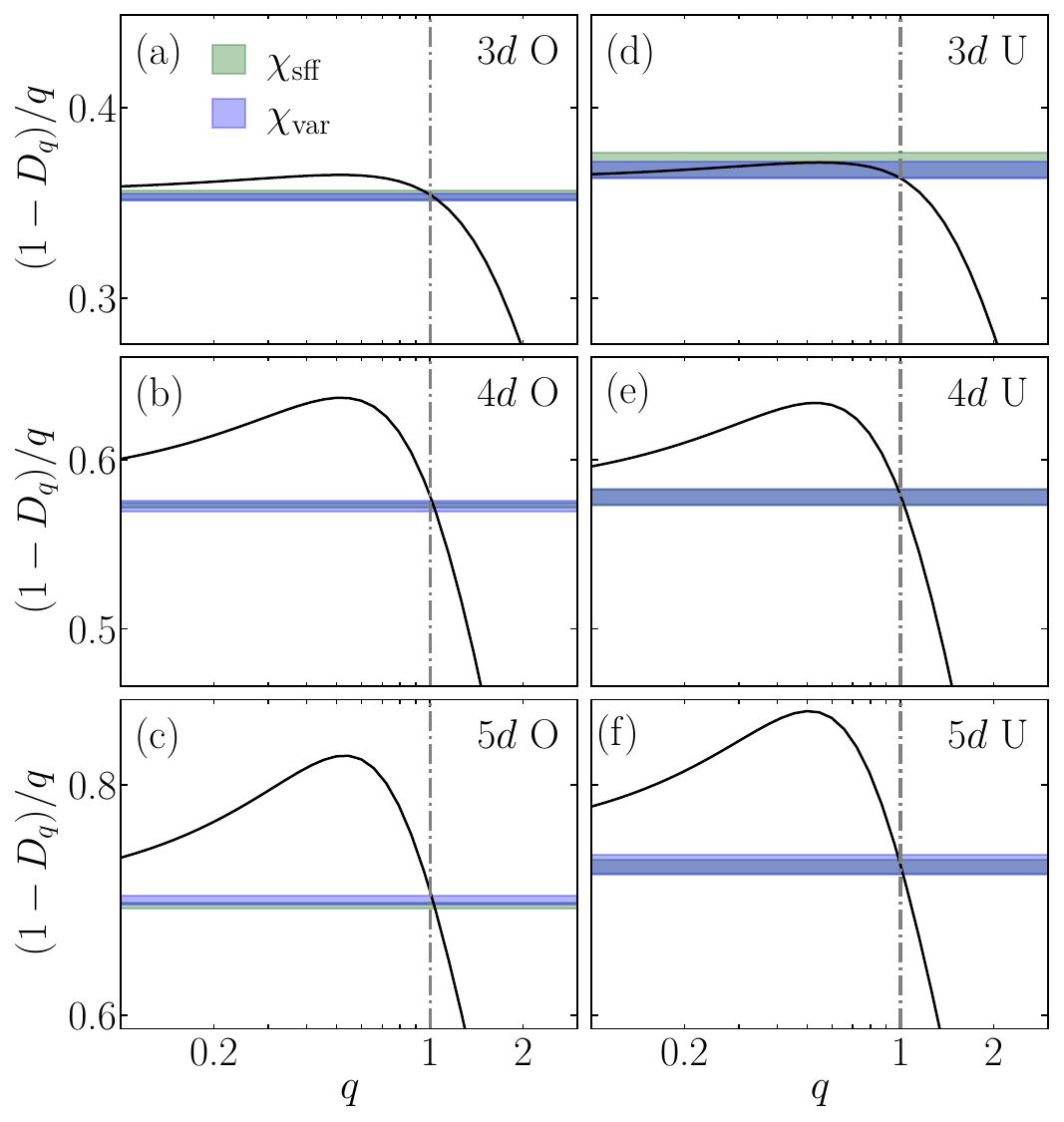}
\caption{
Relation between the scaled fractal dimension $(1-D_q)/q$ (solid lines) and spectral compressibility $\chi$ (horizontal lines of finite width) versus $q$ in the Anderson models of three, four and five dimensions, belonging (a-c) to the orthogonal (O), and (d-f) to the unitary (U) universality class at the critical point. Here the Anderson models of unitary universality class correspond to the definition \eqref{eq:def_hof_couplings} for three and four spatial dimensions (d-e), and to \eqref{eq:def_randphase} for five dimensions (f).
Spectral compressibilities are obtained for the largest available system size via the level number variance, \eqref{eq:def_rel_level_var}, and the SFF, \eqref{eq:def_SFF}, labeled as $\chi_{\rm var}$ and $\chi_{\rm sff}$, respectively. The width of the horizontal lines denotes the estimate of their error bars.
}
\label{fig:fig2}
\end{figure}

We now test the connection between different fractal dimensions $D_q$ and the spectral compressibility $\chi$ in the various Anderson models, extracting $\chi$ for the largest available system size at the critical point using both the level number variance and the SFF.
In Fig.~\ref{fig:fig2} we plot the curves for $(1-D_q)/q$ versus $q$, and we find that the curves intersect with the horizontal line representing $\chi$ at the values of $q$ that are to high numerical precision given by $q=1$.
This provides strong evidence for the validity of the relation in Eq.~(\ref{eq:relation}) for all the Anderson models under consideration, both in the orthogonal and the unitary  case, and for all considered spatial dimensions.

\begin{figure}
\centering
\includegraphics[width=\columnwidth]{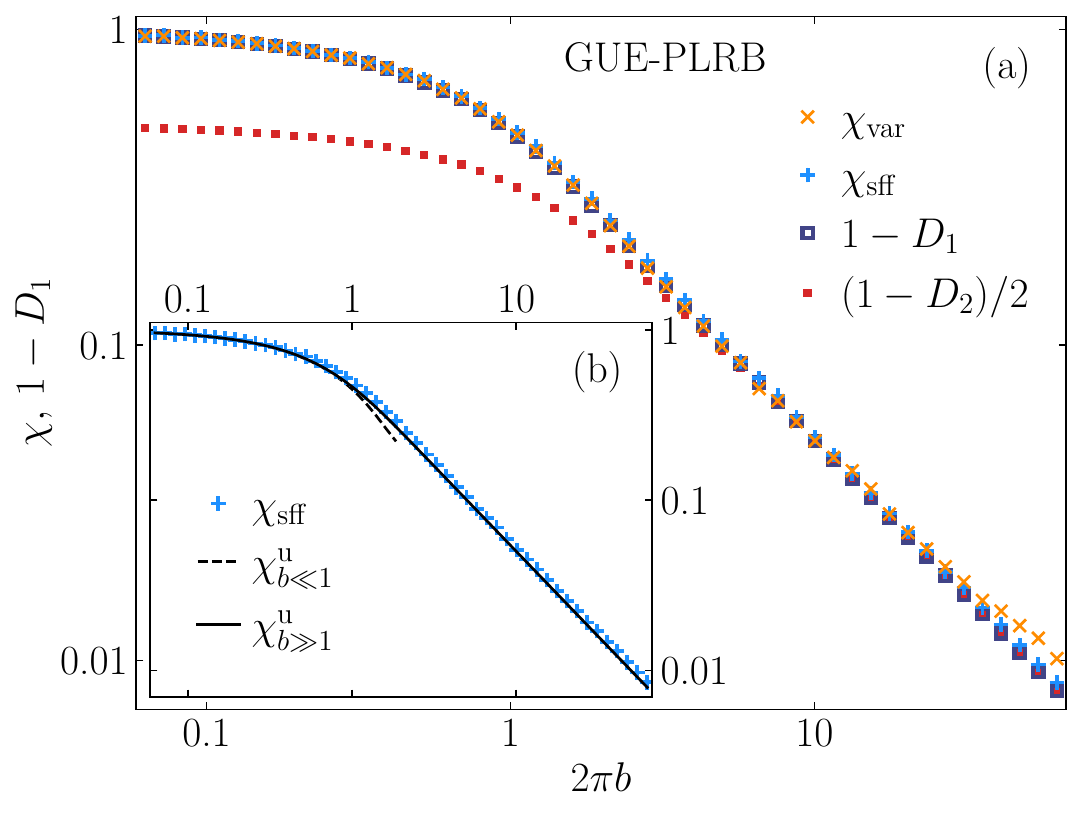}
\caption{
(a) Relation between the spectral compressibility $\chi$ (obtained via the level number variance, $\chi_{\rm var}$, and the SFF, $\chi_{\rm sff}$) and the subtracted fractal dimension
$1-D_1$, at the critical manifold of the GUE-PLRB model, versus $2\pi b$.
We also plot $(1-D_2)/2$ versus $2\pi b$, matching the spectral compressibility only in the weak multifractality limit $2\pi b \gg 1$. 
(b) Comparison of the numerically obtained $\chi_{\rm sff}$ and the theoretical predictions $\chi_{b\ll 1}^{\rm u}$ and $\chi_{b\gg 1}^{\rm u}$, given by Eqs.~(\ref{eq:chi-unit_smallb}) and~(\ref{eq:ans-chi-unit}), respectively.
}
\label{fig:fig3}
\end{figure}

We then extend our analysis to the PLRB models of orthogonal and unitary symmetry class in the broad regime of the tuning parameter $b$.
Results in Fig.~\ref{fig:fig3}(a) for the GUE-PLRB model, and in \textit{Materials and Methods} for the GOE-PLRB model, give rise to the accurate agreement between $1-D_1$ and $\chi$ for all $b$.
We also note that the more general relation $\chi = (1-D_q)/q$ holds true in the weak multifractality limit $2\pi b \gg 1$, as exemplarily illustrated in Fig.~\ref{fig:fig3}(a) for $q=2$ that corresponds to the original relation from Ref.~\cite{Chalker_Kr_Ler}.

Our thorough numerical analysis hence strongly supports validity of the relation in~\eqref{eq:relation} in both local and non-local models at criticality, irrespective of the symmetry class and lattice dimensionality.

\subsection*{Limiting cases:~Analytical insights}

We corroborate our numerical results with analytical insights in certain limiting cases.
In particular, analytical tools are available for the PLRB models to obtain exact solutions for the spectral compressibility in the limits $2 \pi b\ll 1$ and $2\pi b\gg 1$, corresponding to the limits of strong and weak multifractality, respectively.
In the limit of strong multifractality, $2\pi b \ll 1$, the PLRB matrices are almost diagonal and  the spectral compressibility was calculated as a perturbative expansion in the number of resonating energy levels up to order $b^2$ for the GUE-PLRB and up to linear in $b$ terms for GOE-PLRB model~\cite{Mirlin_Evers, Yevtushenko2006}\footnote{There was a misprint in Ref.~\cite{Yevtushenko2006}. It is corrected in~\eqref{eq:chi-unit_smallb}.}. The results of this expansion are: 
 \begin{align} \label{eq:chi-unit_smallb}
    \chi^\text{u}(b) &= \chi^\text{u}_{b\ll 1}(b) + O(b^3)\\ \nonumber  
    \chi^\text{u}_{b\ll 1}(b) &= 1 - \sqrt{2}(\pi b) + \frac{4}{\sqrt{3}}(2 - \sqrt{3})(\pi b)^2\;,\\   
    \label{eq:chi-orth-smallb}
    \chi^\text{o}(b) &= \chi^\text{o}_{b\ll 1}(b) + O(b^2)\\ \nonumber
    \chi^\text{o}_{b\ll 1}(b) &= 1 - \sqrt{2} (2 b)\,,
\end{align}
where $\chi^\text{u}$ and $\chi^\text{o}$ correspond to the spectral compressibilities in the GUE-PLRB and GOE-PLRB models, respectively.

\begin{figure}
\centering
\includegraphics[width=\columnwidth]{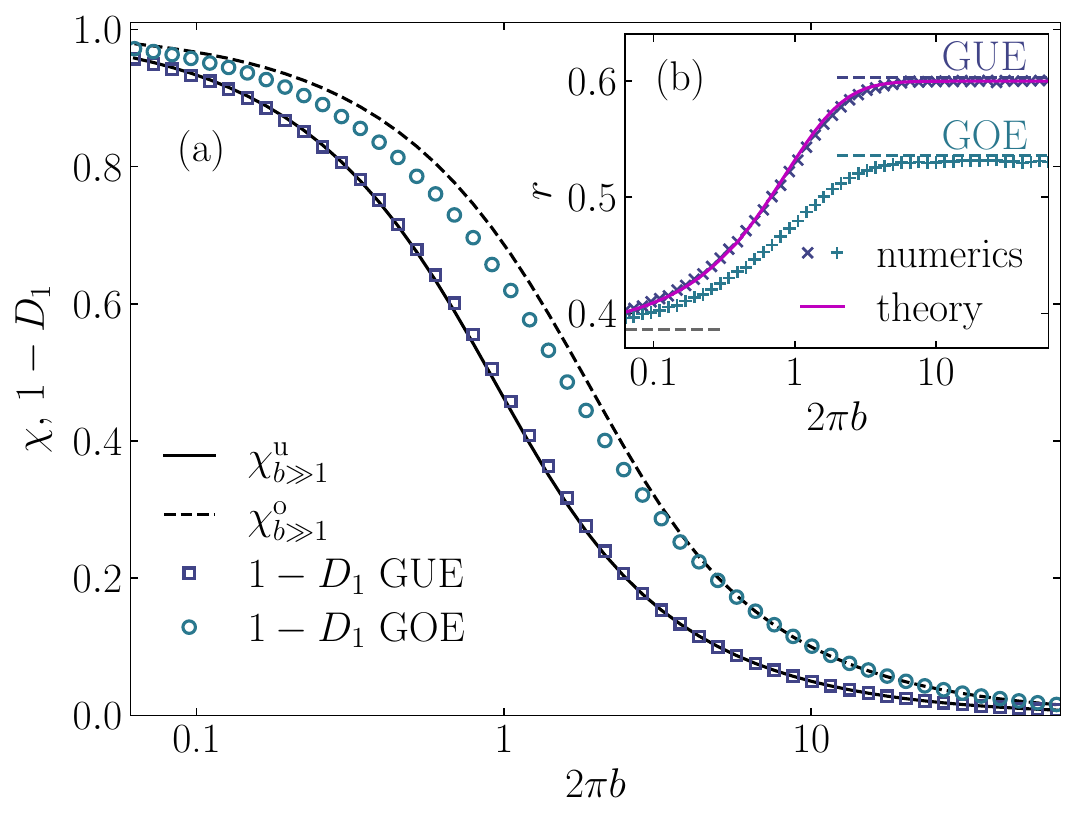}
\caption{
(a) Spectral compressibility $\chi$ and the subtracted fractal dimension $1-D_1$ versus $2\pi b$, at the critical manifold $a=1$ of the PLRB model in the unitary (GUE) and the orthogonal (GOE) class. Numerical results for $1-D_1$ are displayed by symbols and the analytical results for $\chi_{b\gg 1}^{\rm u}$ and $\chi_{b\gg 1}^{\rm o}$, see Eqs.~(\ref{eq:ans-chi-unit}) and~(\ref{eq:ans-chi-orth}), respectively, are displayed as lines.
(b) Average gap ratio $r$ in the PLRB models versus $2\pi b$, at $a=1$.
Symbols are numerical results and the line for the GUE-PLRB model denotes $r$ obtained from the Wigner-Dyson formalism of Ref.~\cite{Bogom_Giraud} with the modified kernel $F(\omega)$ from~\eqref{kernel}.
}
\label{fig:fig4}
\end{figure}

In the limit of weak multifractality, $2\pi b \gg 1$, the PLRB models can be mapped onto the non-linear sigma model~\cite{PLBRM}. Then one can apply the relationship between the classical and quantum spectral determinants~\cite{AA} and compute~\cite{KrMutta, KrTsvel} the density-of-states correlation function $\tilde{K}(\omega)$, which can be thought of as the Fourier transform of the SFF $K(\tau)$, see {\it Materials and Methods}.
Integrating $\tilde{K}(\omega)$ over $\omega$, one obtains, according to~\eqref{chi-SFF}, the spectral compressibility $\chi$ for the unitary and the orthogonal ensembles, respectively,
\begin{align}
\label{eq:ans-chi-unit}
    \chi_{b\gg1}^\text{u}(b) &= \frac{1}{4\pi b} + 1 - \coth (4 \pi b)\;,\\
    \label{eq:ans-chi-orth}
    \chi_{b\gg1}^\text{o}(b) &= \frac{1}{2\pi b} + 1 - \coth (2 \pi b)\,.
\end{align}
Importantly, Eqs.~(\ref{eq:ans-chi-unit}) and~(\ref{eq:ans-chi-orth}) can also be obtained from the Wigner-Dyson formalism~\cite{Mehta} with the modified kernel,
\begin{equation}\label{kernel}
F(x-y)=\frac{\sin(\pi (x-y))}{4b\,\sinh(\pi (x-y)/4b)}\,,
\end{equation}
see {\it Materials and Methods} for further details.
This kernel corresponds to the position statistics of 1$d$ free fermions at a finite temperature $T=1/4b$. In particular, the density correlation function is given by $1-F(x-y)^{2}$, like the density of levels correlation function in the GUE-PLRB model. This establishes a correspondence between the critical level statistics and  the position statistics of 1$d$ fermions at a finite temperature. At weak multifractality (corresponding to small temperature), this analogy holds for all three Dyson universality classes \cite{KrTsvel}.  

In passing, we note that also the fractal dimensions $D_q$ in the critical PLRB models were obtained analytically up to the first order in $b$ and $b^{-1}$ for the limiting cases of strong and weak multifractality, respectively~\cite{Mirlin_Evers, Bogomol}.
In both limits the relation $\chi + D_1 =1$ was thereby analytically confirmed up to first order. However, no analytical predictions are available for $D_1$ in the intermediate regime, which is the most relevant for physical systems.

\subsection*{Towards a general relationship}

Surprisingly, we find that further steps can be made connecting the closed-form expressions in the limiting cases to the non-perturbative regime.
Specifically, we observe in the GUE-PLRB model the almost perfect agreement between the analytical prediction for $\chi_{b\gg 1}^{\rm u}$ from~\eqref{eq:ans-chi-unit} and the numerical results for the spectral compressibility, $\chi_\text{sff}$, and hence with the subtracted fractal dimension, $1-D_1$, in a broad regime of the parameter $b$.
These results are shown on a log-log scale in Fig.~\ref{fig:fig3}(b) and on a linear-log scale in Fig.~\ref{fig:fig4}(a).

The accuracy of our surmise is remarkable provided that by derivation,~\eqref{eq:ans-chi-unit} is only valid for $b\gg 1$ where the non-linear sigma-model applies, see the discussion in Ref.~\cite{KrTsvel}.
Moreover,~\eqref{eq:ans-chi-unit} does not have exactly the same  expansion in powers of $b$ as the exact expansion in the opposite limit $b\ll 1$, given by~\eqref{eq:chi-unit_smallb}.
This means that the accuracy of our surmise in~\eqref{eq:ans-chi-unit} is similar to the accuracy of the celebrated Wigner surmise $P_\text{W}(s)$ for the distribution $P(s)$ of closest gaps $s$, i.e., $P_\text{W}(s)=32 (s/\pi)^2 {\rm exp}[-4 s^2/\pi]$. While $P_\text{W}(s)$ is very close to the exact gap distribution given by the corresponding Fredholm determinant~\cite{Mehta}, it has a different coefficient in front of $s^{2}$ at small $s$ and a different coefficient in the Gaussian behavior at large $s$.

It is also known that the corresponding Wigner surmise for the GOE has more significant deviations from the exact $P(s)$ than that for the GUE. The same is true for $\chi_{b\gg 1}^{\rm o}$ from~\eqref{eq:ans-chi-orth}, which is in Fig.~\ref{fig:fig4}(a) compared to the exact numerical results for the GOE-PLRB model. While the agreement between the latter and $\chi_{b\gg 1}^{\rm o}$ is perfect at large $2\pi b$, the deviations become significant at intermediate values of $2\pi b$.

\begin{figure}
\centering
\includegraphics[width=\columnwidth]{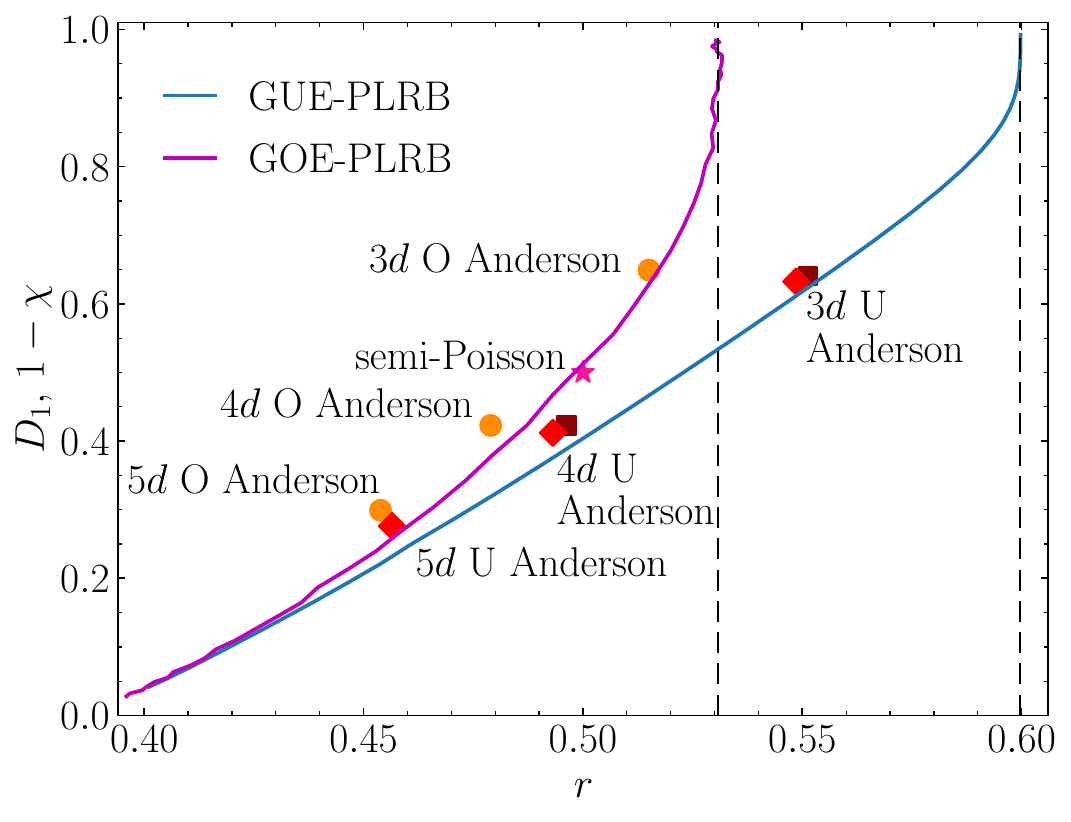}
\caption{
Universal functions $D_{1}( r)$ for the critical models of the unitary (blue curve) and orthogonal (magenta curve) symmetry class. They were derived analytically for the GUE-PLRB model as explained in the text, and numerically for the GOE-PLRB model.
Symbols: numerical results for different critical models (points with the statistical error bars corresponding to their sizes): 3$d$, 4$d$ and 5$d$ Anderson models of orthogonal (dots) and unitary symmetry class (diamonds for the model in~\eqref{eq:def_randphase}, squares for the model in~\eqref{eq:def_hof_couplings}).
The star corresponds to systems with the "semi-Poisson" distribution of gaps where $D_{1}= r =1/2$~\cite{Bogo2020, Bogo_Giraud2021}.
The vertical dashed lines are the Wigner-Dyson limits $r=0.5997$ and $r=0.5307$.
}
\label{fig:fig5}
\end{figure}

To further corroborate that the analogy with the free fermions at a finite temperature gives a very accurate surmise for different spectral statistics in the GUE case, we compute the average gap ratio $r$ by the Fredholm determinant formalism~\cite{Mehta,Bogom_Giraud} based on the same modified kernel, Eq.~(\ref{kernel}), as the one that was employed to compute~\eqref{eq:ans-chi-unit}. Again we find an almost perfect agreement with the numerically obtained $r$ across the whole critical manifold, see Fig.~\ref{fig:fig4}(b). The details of this semi-analytical calculation will be published elsewhere.
We hence conclude that the Wigner-Dyson theory based on a modified kernel,~\eqref{kernel},
gives a very accurate prediction for different spectral statistics of the GUE-PLRB model in the entire critical manifold.

Finally, we derive a relationship between the fractal dimension $D_{1}$ and the average gap ratio
$r$, which allows us to demonstrate its universality for different critical models, both local and non-local, with and without time-reversal symmetry.
The derivation is carried out semi-analytically for the GUE-PLRB model, combining~\eqref{eq:ans-chi-unit}
with the results for $r$ obtained from the Wigner-Dyson formalism with the modified kernel,~\eqref{kernel}.
Indeed, the functions $\chi(b)$ and $ r(b)$ parametrically define the function $\chi( r)$. Then, using the relation $1-D_{1} = \chi$ from~\eqref{eq:relation}, we can express $D_{1}$ in terms of $ r$. The same operation for the GOE-PLRB model is done using numerical results for $\chi(b)$ and $ r(b)$. The results are shown in Fig.~\ref{fig:fig5}.

We test the universality of the functions $D_{1}(r)$ by calculating the values of $D_{1}$ and $ r$ at the critical points of the Anderson models in 3$d$, 4$d$ and 5$d$, both in the orthogonal and unitary symmetry class, as well as by adding the point $D_{1}=r=1/2$ that characterizes systems with semi-Poisson statistics, see, e.g., Refs.~\cite{Bogo2020,Bogo_Giraud2021}. 
The results (symbols) in Fig.~\ref{fig:fig5} show a reasonably good agreement with our predictions (solid lines).
Nevertheless, the agreement becomes less perfect in higher-dimensional Anderson models.
We attribute these deviations to systematic errors that emerge in numerical calculations for finite systems, in which the accessible linear dimensions $L=\sqrt[\leftroot{1}\uproot{1}d]{N}$ decrease with the increase of dimensionality.
We note that the dependence of $D_{1}(L)$ near criticality typically exhibits a shallow minimum~\cite{Skvortsov16, AltKrIoffe16, ScardVanoni, ScardVanoni_high_d}. Hence one may observe a decrease of $D_{1}(L)$ at the accessible sizes $L$, while at even larger sizes, $D_{1}(L)$ may start to grow towards the ergodic limit $D_{1}=1$.
Therefore the numerical results at limited system sizes may suggest the system's tendency towards localization, while it is actually still in the delocalized phase.
Consequently, underestimating the critical value $W_{c}$ gives rise to a systematic overestimate of $D_{1}(W_c)$, which is consistent with the observations in Fig.~\ref{fig:fig5}.

Using the analytically derived universal function $D_{1}(r)$ one may also predict the values of $r$ for another long-disputed transition, namely, the integer quantum Hall effect transition at the center of the Landau band.
Carrying out a high-precision computation on the Chalker-Coddington network~\cite{Evers_D1}, the value of $D_{1}$ for this transition was obtained to be $D_1 = 0.8702$, and a field-theoretical calculation by Zirnbauer~\cite{Zirnbauer2019} predicted the critical value $D_{1}=7/8$.
The (unitary) universal function derived here gives $r=0.5962$ for $D_{1}=0.8702$ and $r=0.5966$ for $D_{1}=7/8$. 
In future, it would be interesting to compare this prediction with the numerical results on the Chalker-Coddington networks.

\section*{Discussion}

Our work establishes generality of the relationship between the spectral compressibility and the wavefunction fractal dimension, $\chi + D_1 = 1$ from~\eqref{eq:relation}, for physical models in various dimensions across different symmetry classes.
Moreover, we also found indications of relationships between the spectral and wavefunction properties beyond~\eqref{eq:relation}, which is triggered by the relevance of the Wigner-Dyson formalism with the modified kernel in a broad regime of the critical manifold.
This opens doors for further exploration of spectral properties beyond compressibility, such as the average gap ratio.

Our results also raise several outstanding questions for future research.
One of them is the strict definition, and possibly the uniqueness, of the preferred basis in which the wavefunction fractal dimension takes a minimal value. By definition, the preferred basis should be independent of the Hamiltonian realization and thus the energy basis is excluded.

The preferred basis of the models studied here appears to coincide with the "natural" basis in which the model is formulated.
This is certainly not always the case, and a counterexample may be given by the ensemble of Toeplitz matrices from Refs.~\cite{Bogo2020, Bogo_Giraud2021}.
These matrices are formulated in the basis where their entries $H_{nm}=f(|n-m|)$, with $f(n)$ being the i.i.d.~Gaussian random variables. However, the preferred basis for these models is the momentum basis, in which eigenfunctions show multifractal statistics. 

Further, one can also imagine a continuous family of preferred bases with the same minimal $D_{1}$, such as a "mexican hat" of preferred bases.
In this case, the averaging over such bases does not eliminate the deviation of spectral statistics from the Wigner-Dyson statistics, as demonstrated in Ref.~\cite{MNS}.

All critical systems studied here contain parameters of the Hamiltonians that are independent of the system size, and the local spectrum in the preferred basis is a random Cantor set~\cite{Alt_Kr_Cantor}. These appear to be sufficient conditions for \eqref{eq:relation} to hold. In such systems the local density-of-states correlation function  is a power law, with an exponent that belongs to the interval $(0,1)$ for all energies down to the level spacing~\cite{Alt_Kr_Cantor}, thus connecting the large energy scale properties (spectral compressibility) and the small energy scale properties, such as the average gap ratio.

Another outstanding question is the generalization of our results to interacting systems, and their possible relevance for quantum dynamics. From the perspective of spectral properties in interacting systems, recent work has suggested that the dynamics of the spectral form factor may provide information about the many-body spectral compressibility~\cite{suntajs_vidmar_22, Hopjan23b}. Information about the fractal dimension $D_2$ of the many-body wavefunctions can also be extracted from the power-law decay of survival probability of many-body wavefunctions~\cite{hopjan2023}.

However, for many-body systems a "natural" basis is not obvious and the problem of a preferred basis is highly non-trivial. A recent study \cite{Mace2019} demonstrated that the fractal dimensions $D_{1}$ and $D_{2}$ are different for a spin-1/2 XXZ chain in the configurational (bitstring) basis, and in the Fock basis built by Anderson orbitals, taking smaller values in the latter. This means that the Fock basis is closer to the preferred one than the configurational basis. Some recent attempts to approach the problem of preferred basis in many-body systems are described in Refs.~\cite{fin1,fin2}.

Establishing possible relationships between spectral and wavefunction features in interacting quantum systems may provide new impetus to the analysis of critical properties at the many-body ergodicity breaking transitions, including the many-body localization transition that has recently experienced a lively discussion about the position of the transition point in certain paradigmatic models~\cite{Sierant_lewenstein_25}.

\matmethods{

\subsection*{Anderson models in unitary universality class}

We consider two different instances of Anderson models belonging to the unitary universality class.
The first, corresponding to random hopping phases, was defined in~\eqref{eq:def_randphase} in the main text.
For the second option we dress the hopping term with a complex phase in $d-2$ dimensions according to 
\begin{align}
    t_{\bm{r},\bm{r}+\bm{x}} &\equiv t\,, & \hspace*{0.0cm} & t_{\bm{r},\bm{r}+\bm{z}}\equiv t\, \text{e}^{2\pi i \phi x}\,, \nonumber  \\
    t_{\bm{r},\bm{r}+\bm{y}} &\equiv t\,, & \hspace*{0.0cm} & t_{\bm{r},\bm{r}+\bm{w}} \equiv t\, \text{e}^{2\pi i \phi (x+y)}\,, \label{eq:def_hof_couplings}
\end{align}
where the fourth spatial direction $\bm{w}$ is only present in the case of four dimensions. In the 3$d$ case, this definition corresponds to the presence of a magnetic field of flux $\phi$ in $\bm{y}$ direction. We choose $\phi=1/4$ throughout this study. For this strength of the magnetic field a transition point of $W_c=18.38$ was found in a previous study~\cite{Slevin_Ohtsuki_1997}. For the 4$d$ case no previous studies of the transition are available, although the coupling of fluxes as in~\eqref{eq:def_hof_couplings} was discussed in the context of topological insulators~\cite{Mochol-Grzelak_2019}. Therefore we carry out a scaling analysis to extract the transition point, yielding $W_c=37.1$, see the text below for further details. Due to the limited available system sizes in higher dimensions we use both periodic and anti-periodic boundary conditions in the $4d$ case, while we refrain form a further generalization of this model to 5$d$, since the accessible system sizes for which periodic or anti-periodic boundary conditions can be chosen are limited.

\subsection*{Results for GOE-PLRB model}

We repeat the same analysis, as shown in Fig.~\ref{fig:fig3}, for the PLRB model of orthogonal universality class. The results are displayed in Fig.~\ref{fig:fig_plrb_goe_appndx}. Again we observe validity of the relation $\chi + D_1 = 1 $ across the whole critical manifold.

\begin{figure}
\centering
\includegraphics[width=\columnwidth]{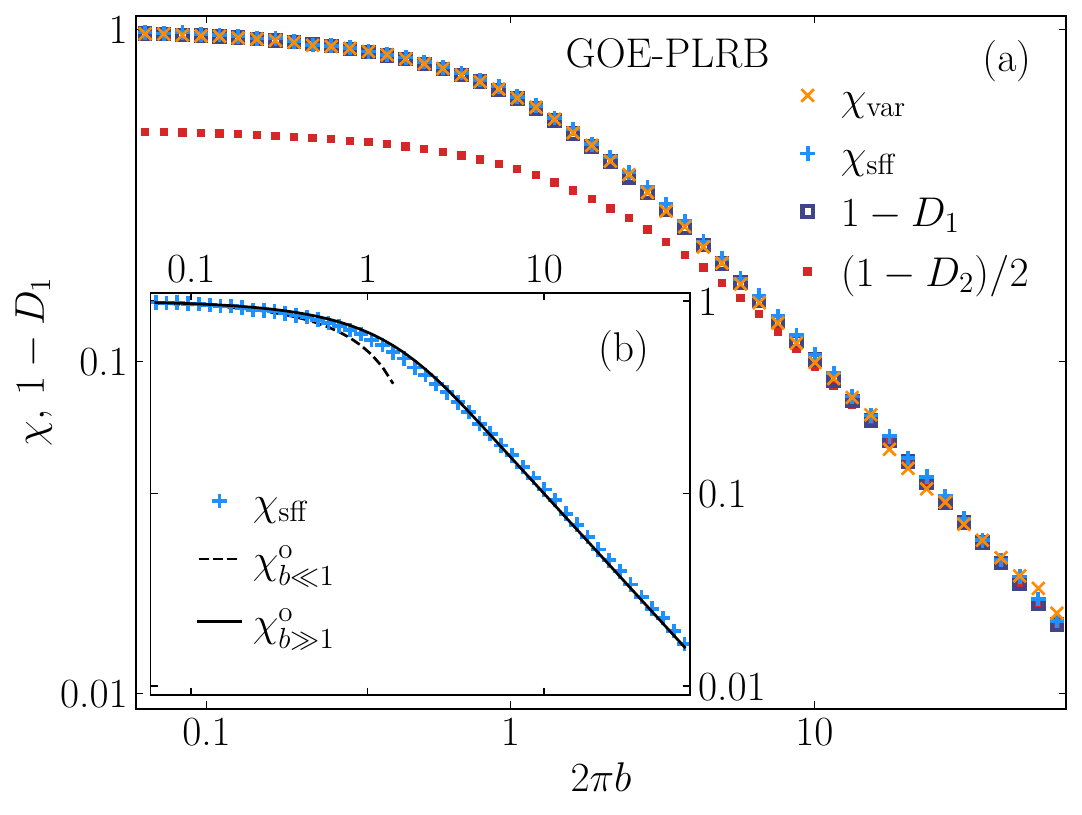}
\caption{
(a) Relation between the spectral compressibility $\chi$ (obtained via the level number variance, $\chi_{\rm var}$, and the SFF, $\chi_{\rm sff}$) and the subtracted fractal dimension
$1-D_1$, at the critical manifold of the GOE-PLRB model, versus $2\pi b$.
We also plot $(1-D_2)/2$ versus $2\pi b$, matching the spectral compressibility only in the weak multifractality limit $2\pi b \gg 1$. 
(b) Comparison of the numerically obtained $\chi_{\rm sff}$ and the theoretical predictions $\chi_{b\ll 1}^{\rm o}$ and $\chi_{b\gg 1}^{\rm o}$, given by~\eqref{eq:chi-orth-smallb} and~\eqref{eq:ans-chi-orth}, respectively.
}
\label{fig:fig_plrb_goe_appndx}
\end{figure}

\subsection*{Determination of transition points}

While the transition points of the Anderson models in the orthogonal symmetry class, as well as in certain lattice geometries in the unitary symmetry class, are known to high accuracy~\cite{Slevin}, we are not aware of any comparable analyses of the 4$d$ unitary Anderson model described by~\eqref{eq:def_hof_couplings}, and the 5$d$ unitary Anderson model with random hopping phases from~\eqref{eq:def_randphase}.
Therefore, we carry out a scaling analysis to extract the transition point in these models by applying the cost function minimization approach using the average gap ratio $r$, defined in~\eqref{eq:def_gap_ratio}, as the transition indicator.
Details of the method can be found in Refs.~\cite{Suntajs_2020, Suntajs_2020_2}.
The results for $r$ and the corresponding optimal scaling solutions are shown in Fig.~\ref{fig:trans_points_unitary_And}. We find the transition point at $W_c\approx37.1$ for the 4$d$ unitary Anderson model from~\eqref{eq:def_hof_couplings} and at $W_c\approx61.3$ for the 5$d$ unitary Anderson model from~\eqref{eq:def_randphase}.

\begin{figure}
\centering
\includegraphics[width=\columnwidth]{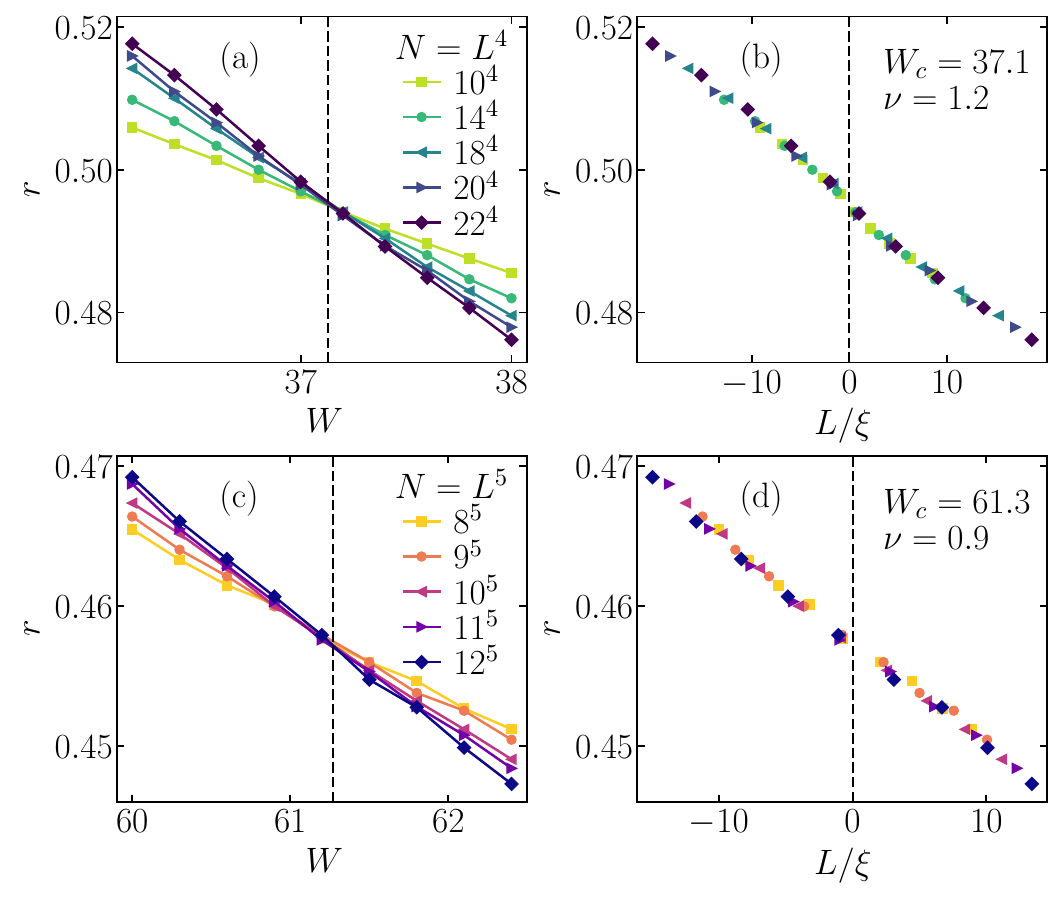}
\caption{
Determination of the transition points for the unitary Anderson models using the average gap ratio $r$, see~\eqref{eq:def_gap_ratio}, as the transition indicator.
(a, b) 4$d$ unitary Anderson Model from~\eqref{eq:def_hof_couplings}.
(c, d) 5$d$ unitary Anderson Model from~\eqref{eq:def_randphase}.
Panels (a) and (c) show the results for $r$ versus $W$ for different total system sizes $N=L^d$.
In panels (b) and (d) we show the optimal scaling collapse of $r$ versus $L/\xi$, where $L$ denotes the linear size and $\xi=\mathrm{sign}(W-W_c)\,|W-W_c|^{-\nu}$.
The optimal scaling collapse is obtained using the cost function minimization approach described in Refs.~\cite{Suntajs_2020,Suntajs_2020_2}. 
Vertical dashed lines in all panels correspond to the transition point $W=W_c$, and the extracted values of $W_c$ and $\nu$ are given in panels (b) and (d).
}
\label{fig:trans_points_unitary_And}
\end{figure}

\subsection*{Details of numerical calculations}
\label{subsec:numerical_details}
Here we provide details about the numerical algorithms we used, the averaging over eigenstates and Hamiltonian realizations, the spectral unfolding, and the two different methods to extract the spectral compressibility, i.e., the level number variance and the spectral form factor. \\
\textbf{Numerical methods.}
The calculation of level number variance and spectral form factor requires access to a great fraction, respectively all of the eigenvalue spectrum for the model under investigation. Therefore we use exact diagonalization routines for this purpose. To calculate the average gap ratio, as well as the fractal dimensions, it suffices to determine a finite number of eigenstates in the middle of the spectrum. To this end we use the method of Polynomial-filtered Exact Diagonalization~\cite{sierant_lewenstein_2020}, allowing us to reach large system sizes for the Anderson Models. Since this method requires sparsity of the Hamiltonian matrix, it is not applicable to the dense PLRB models, for which we resort to exact diagonalization instead. \\
\textbf{Averaging over eigenstates and Hamiltonian realizations.}
In general, for a single Hamiltonian realization, the averages are carried out over the middle of the spectrum.
The Anderson models exhibit a mobility edge at $W<W_{\rm c}$, i.e., there exist an energy $E_{\rm c}$ that separates localized and extended states.
With increasing disorder strength $W$, the mobility edge moves quickly from the edges of the spectrum to its center, so that almost all the states are either extended or localized, the transition being extremely sharp at $W\approx W_{c}$.
At the transition point $W_{c}$ and close to the mobility edge, there is a divergent critical length $\xi\propto |W-W_{c}|^{-\nu}$ (in the center of the energy band) or $\xi\propto |E-E_{c}|^{-\nu}$ (close to the mobility edge $E_{c}$), such that for $L<\xi$ all the eigenstates are critical and exhibit multifractality.

The maximal detuning from the critical disorder strength $W_c$, or from the mobility edge $E_{c}$, to observe the critical states is  $|1-W/W_{c}|\sim|1-E/E_{c}|\sim L^{-\frac{1}{\nu}}$. In units of the mean level spacing, $\delta\propto 1/N$, it is therefore proportional to $L^{d-\frac{1}{\nu}}\gtrsim L^{\frac{1}{\nu}}$, which,  according to the Harris criterion $d\nu>2$~\cite{Harris}, is a macroscopically large number in the large-$L$ limit. So, despite the spectral window of the critical states shrinking to zero in the large-$L$ limit, the number of levels in this window goes to infinity, allowing to define the critical spectral and eigenfunction statistics.

\begin{table}
\def\arraystretch{1.2}
\caption{Details of numerical simulations}
\label{tab:real_nums}
\resizebox{\columnwidth}{!}{%
\begin{tabular}{c|c|c|c|c}
Model       & Quantity   & system sizes $N$                                                                                            & \begin{tabular}[c]{@{}c@{}}$n_R^{*}$\end{tabular} & \begin{tabular}[c]{@{}c@{}}$n_S^{**}$\end{tabular} \\ \hline
3$d$ And & $\chi$     & $32^3$, $36^3$, $40^3$                                                                                      & 1000                                                                          &                                                                                 \\
            & $r,\, D_1$ & $24^3$, $32^3$, $40^3$, $48^3$, $56^3$, $64^3$                                                              & 2000                                                                          & 500                                                                             \\ \hline
4$d$ And & $\chi$     & $12^4$, $14^4$, $16^4$                                                                                      & 2000                                                                          &                                                                                 \\
            & $r,\, D_1$ & $12^4$, $14^4$, $16^4$, $18^4$, $20^4$, $22^4$                                                              & 5000                                                                          & 500                                                                             \\ \hline
5$d$ And & $\chi$     & $7^5$, $8^5$, $9^5$                                                                                         & 2000                                                                          &                                                                                 \\
            & $r,\, D_1$ & $8^5$, $9^5$, $10^5$, $11^5$, $12^5$, $13^5$                                                                & 5000                                                                          & 500                                                                             \\ \hline
PLRB        & $\chi$     & \begin{tabular}[c]{@{}c@{}}$2$k, $4$k, $6$k, $8$k, \\ $10$k, $12$k, $14$k, $16$k, $18$k, $20$k\end{tabular} & \begin{tabular}[c]{@{}c@{}}1000\\ 500\end{tabular}                            &                                                                                 \\
            & $r,\, D_1$ & \begin{tabular}[c]{@{}c@{}}$2$k, $4$k, $6$k, $8$k, \\ $10$k, $12$k, $14$k, $16$k, $18$k, $20$k\end{tabular} & \begin{tabular}[c]{@{}c@{}}1000\\ 500\end{tabular}                            & 10\% of $N$                                                                    
\end{tabular}%
}
$^*$number of different Hamiltonian realizations, $^{**}$number of considered eigenstates per Hamiltonian realization
\end{table}

The actual number of Hamiltonian realizations we average over and the system sizes for all the investigated models can be seen in Tab.~\ref{tab:real_nums}. We use the same system sizes and number of Hamiltonian realizations for the case of orthogonal and unitary universality class.\\
\textbf{Unfolding.}
Prior to calculating the spectral form factor or the level number variance, we set the local mean level spacing to unity at all energy densities by unfolding the spectrum.
To this end, starting from the exact eigenvalues $\varepsilon_\alpha$ for a given Hamiltonian realization, we construct the step function $G(\varepsilon) = \sum_{\alpha = 1}^N \Theta(\varepsilon - \varepsilon_\alpha)$ and fit it with a polynomial of low order $n$ ($n=3$ for Anderson models and $n=5$ for the PLRB models).
The resulting fitting function $g_n(\varepsilon)$ yields the unfolded eigenvalues as $E_\alpha = g_n(\varepsilon_\alpha)$.\\
\textbf{Details of spectral form factor calculations.}
To reduce contributions from the spectral edges, we introduced a Gaussian filtering factor $\eta(E)$ in the definition of the spectral form factor, \eqref{eq:def_SFF}. For a given Hamiltonian realization yielding the unfolded eigenvalues $E_\alpha$, the filtering function is defined as
\begin{align}
    \eta(E) = \exp\left(-\frac{(E_\alpha - \Bar{E})^2}{2\,\eta_0\,\Gamma^2}\right)\, ,
\end{align}
where $\Bar{E}$ and $\Gamma^2$ denote the mean energy, respectively the energy variance, for the given Hamiltonian realization. The factor $\eta_0$ affects the width of the Gaussian and thereby the effective filtering strength. A priori, there is no best choice for $\eta_0$ since this would require perfect knowledge about the mobility edge and boundary effects for a given finite system. Empirically we find that for the Anderson models a filter using $\eta_0\in[0.3,0.7]$ yields the broadest plateau at the transition point, while for the PLRB models a filter using $\eta_0=0.3$ poses a suitable choice. The error estimate provided in Fig.~\ref{fig:fig2} for the plateau value of the spectral form factor, and hence for the spectral compressibility, is based on a scan through the different filtering function widths and slightly different conceivable placements of the time-interval to extract the plateau value.

\textbf{Details of level number variance calculations.}
In our numerical calculations of the energy-dependent spectral compressibility via the level variance, as defined in \eqref{eq:def_rel_level_var}, we proceed in three steps. Starting from the unfolded spectrum for a given Hamiltonian realization, we first cut away a fraction of eigenvalues at the boundaries. We find that for the Anderson models, larger parts of the spectrum have to be considered to optimally unveil the plateau, corresponding to a cutoff of 10\% to 25\% of states at both ends of the spectrum, whereas for the PLRB models, a more narrow focus on only 20\% of states in the middle of the spectrum works best. 

In the next step, after the cutoff of spectral edges, we draw one box of width $\Delta$ in units of mean level spacing $\delta$ ($\sim 1$ after unfolding) for each Hamiltonian realization and place it within the remaining middle part of the spectrum. Empirically we find that it is beneficial to vary the midpoints from realization to realization by random displacements. 
In the Anderson models, in contrast to the PLRB models, a higher degree of randomness for the placement of midpoints is necessary to unveil the plateau optimally. Having obtained the number of levels within the box for all realizations, we calculate their variance and divide by the mean number of levels. Here we want to note that, even for the unfolded spectrum it proofs beneficial to divide by the mean number of levels across al realizations instead of the interval width. We argue that this is caused by the density of states, which is not perfectly flat even after the unfolding procedure. 

In the last step, we repeat the procedure 20 times for differently chosen interval positions in each realization around the middle of the spectrum and average over the resulting spectral compressibility, to reduce fluctuations.

The error estimate provided in Fig.~\ref{fig:fig2} for the spectral compressibility, obtained from the level number variance, is based on a scan through different sizes of the initial boundary cutoff, the amount of randomness in the placement of interval midpoints, and slightly different conceivable intervals to extract the plateau value.

\subsection*{Gap ratio}
A useful measure to characterize the localization transitions is the average gap ratio $r$, defined as~\cite{Huse_Oganes}
\begin{align}
\label{eq:def_gap_ratio}
\begin{split}
    r &= \braket{\Bar{r}}_H \;,\;\;\;
    \Bar{r} = \braket{r_\alpha}_{\alpha}\;, \\
    r_\alpha &= \frac{\min(\varepsilon_{\alpha+1} - \varepsilon_\alpha, \varepsilon_\alpha - \varepsilon_{\alpha-1})}{\max(\varepsilon_{\alpha+1} - \varepsilon_\alpha, \varepsilon_\alpha - \varepsilon_{\alpha-1})} \;,
\end{split}
\end{align}
where  $\braket{\dots}_H$ and $\braket{\dots}_{\alpha}$ denote averaging over different Hamiltonian realizations and the states around the middle of the spectrum, respectively, and the $\varepsilon_\alpha$ denote the exact Hamiltonian eigenvalues. We note that for the gap ratio, we do not require unfolded spectra since the gap ratio is insensitive to the density of states.

For the Wigner-Dyson GOE and GUE random matrix ensembles, the distribution of $r_\alpha$ and its average value $r$ can be exactly computed for any $N$ by the Fredholm determinant formalism described in Refs.~\cite{Bogom_Giraud, Mehta}.
In Ref.~\cite{Bogom_Giraud} a simple and rather precise surmise was derived using a simplified  approach of considering only $3\times 3$ matrix Hamiltonian. Unfortunately, this simplified approach is only good for the Wigner-Dyson kernel,~\eqref{kernel}, at $b\rightarrow\infty$, as it is intimately related with the incompressibility of the system's levels, $\chi=0$. 

In order to analytically compute the average gap ratio $r$ in the limit $N\rightarrow\infty$ for an arbitrary $b$ in the kernel from~\eqref{kernel}, we employed the exact Fredholm determinant formalism as described in Refs.~\cite{Mehta, Bogom_Giraud}.
The result of the calculation of the function $r(b)$ for the GUE-PLRB model is presented in Fig.~\ref{fig:fig4}(b). Remarkably, it matches perfectly with the corresponding function found by exact diagonalization.

\subsection*{Relation between the global density of states correlation function and the spectral form factor}
Consider the spectral form-factor:
\begin{eqnarray}
\label{eq:def_SFF_appx}
    && K(\tau) = \frac{1}{Z}\left\langle \sum_{\alpha,\beta=1}^N \eta(E_\alpha)\eta(E_\beta)\text{e}^{-2\pi i (E_\alpha-E_\beta) \tau} \right\rangle_{\hspace*{-1.2ex}H}\\
    \nonumber && =\frac{1}{Z}  \int d\omega\,e^{2\pi i \omega\tau} \int dE\,\eta(E)\eta(E-\omega)\,  \left\langle  \rho(E) \rho(E-\omega)\right\rangle_{H},    
\end{eqnarray}
where all energies are measured in units of the mean level spacing and
\begin{align}
\rho(E)=\sum_{\alpha=1}^N\delta(E-E_\alpha)
\end{align}
is the (global) density of states (DoS).
One can see that $K(\tau)$ is the Fourier transform of
\begin{align}\label{eta-G}
\frac{1}{Z}\int dE\,\eta(E)\eta(E-\omega)\,G(\omega,E),    
\end{align}
where $G(\omega,E)$ is the correlation function of the (global) DoS:
\begin{align}\label{G}
G(\omega, E)=\left\langle\sum_{\alpha=1}^N\delta(E-E_\alpha)\sum_{\beta=1}^N\delta(E-\omega-E_\beta)\right\rangle_{\hspace*{-1.2ex}H}. 
\end{align}

Now consider the function $G(\omega,E)$ and its relation with the level number variance $\Sigma^2$ in an energy window $\Delta$ centered at $E=0$. The latter is given by definition
\begin{align}\label{variance1}
\Sigma^2=\left\langle \Bigg\{\int_{-\Delta/2}^{\Delta/2}dE  [\rho(E) -\left\langle\rho(E)\right\rangle_H]\Bigg\}^2 \right\rangle_{\hspace*{-1.2ex}H}.
\end{align} 
Using the definition of the DoS we obtain: 
\begin{align}\label{variance2}
\Sigma^2=\left\langle \Bigg\{\int_{-\Delta/2}^{\Delta/2}dE  \sum_{\alpha=1}^N[\delta(E-E_\alpha) -\left\langle\delta(E-E_\alpha)\right\rangle_H]\Bigg\}^2 \right\rangle_{\hspace*{-1.2ex}H}
\end{align} 
and rewrite as:
\begin{align}\label{variance3}
\Sigma^2=\int_{-\Delta/2}^{\Delta/2}dE\int_{-\Delta/2}^{\Delta/2}dE'\,[G(E-E',E)- \left\langle\rho(E) \right\rangle_{H}^{}\left\langle\rho(E')\right\rangle_{\hspace*{-0.5ex}H}^{}].
\end{align}
At small (compared to the total spectral band-width) width of the filtering function $\eta(E)$ and corresponding small $\Delta$ and  $\omega$ the correlation function $G(\omega, E)$ is  approximately independent of $E$:
\begin{align}
 G(\omega,E)\approx G(\omega).
\end{align}
Similarly, the mean DoS $\rho(E)$ is almost independent of $E$ and can be replaced by $\rho_{0}\equiv\rho(0)=1$.
Integrating over $E+E'$ at a fixed $\omega$  we obtain:
\begin{align}\label{var-R}
 \Sigma^2=\int_{-\Delta}^{\Delta}(\Delta -|\omega|)\,R(\omega)\,d\omega,
\end{align}
where
\begin{align}
R(\omega)=G(\omega) -1.
\end{align}
It immediately follows from Eq.(\ref{var-R}) that:
\begin{align}
\chi(\Delta)=\frac{d\Sigma^2}{d\Delta}=\int_{-\Delta}^{\Delta}R(\omega)\,d\omega.
\end{align}
Notice that $R(\omega)$ in~\eqref{G} contains a $\delta(\omega)$ part coming from the term in the sum with $\alpha=\beta$.

Thus we have
\begin{align}\label{chi-G}
\chi(\Delta)=1+2\int_{+0}^{\Delta}R(\omega)d\omega,
\end{align}
where we assume that all energies are measured in units of $\delta$ and $R(\omega)$ is an even function.
Apart from this last assumption,~\eqref{chi-G} is absolutely general. 

The critical $R(\omega)$ is characterized by the limiting $R_{\infty}(\omega)=\lim_{N\rightarrow\infty}R(\omega)$.  It decrease fast enough as $\omega\rightarrow\infty$, so that the integral in~\eqref{chi-G} is  convergent at $\Delta\rightarrow\infty$ and negative due to level repulsion. In contrast, for an insulator $R_{\infty}(\omega)\equiv 0$ at $N\rightarrow\infty$. For a metal, $R(\omega)$ is described by the Wigner-Dyson theory for $\omega$ much smaller than the Thouless energy $E_{Th}$, which, in units of $\delta$, tends to infinity $E_{Th}/\delta\equiv g(N) \rightarrow\infty$ as $N\rightarrow\infty$. 

If the energy range of a filter function is $N$-independent, it will be infinite  in the limit $N\rightarrow\infty$ when measured in units of $\delta\sim 1/N$. Assuming this limit is taken first and the integral in~\eqref{chi-G} is convergent, we conclude that the critical spectral compressibility tends to a constant $0<\chi<1$ at $\Delta\rightarrow\infty$. 
\begin{align}\label{chi-fin}
\chi=\lim_{\Delta\rightarrow\infty}\lim_{N\rightarrow\infty}\chi(\Delta)=1+2\int_{+0}^{\infty} R_{\infty}(\omega)\, d\omega
\end{align}
For localized states~\eqref{chi-fin} gives $\chi=1$, and for the ergodic delocalized ones $\chi=0$, since for all symmetry classes of Wigner-Dyson theory $\int_{+0}^{\infty}\lim_{N\rightarrow \infty}R(\omega)=-1/2$.

Now return back to~\eqref{eta-G} and cast it as:
\begin{align}\label{FT}
K(\tau)=FT\left\{\frac{1}{Z}\int dE\,\eta(E)\eta(E-\omega)\,[R(\omega)+1]\right\}.
\end{align}
For a Gaussian filter $\int dE \,\eta(E)\eta(E-\omega)$ is a Gaussian function   of the width which is $\sqrt{2}$ larger than that of $\eta(E)$. If this width is $N$-independent in conventional energy units,   it will be infinite when measured in units of $\delta$ in the large-$N$ limit. At the same time, for critical states the range of the correlation function $G(\omega)$  which makes the dominant contribution to~\eqref{chi-fin}, is finite. This separation of scales allows to replace:
\begin{align}
\frac{1}{Z}\int dE\eta(E)\eta(E-\omega)\rightarrow \frac{1}{Z}\int dE [\eta(E)]^{2}=const.
\end{align}
One can always take the normalization $Z$ such that this constant is equal to 1. 
Taking also into account that the Fourier transform of a constant is a $\delta(\tau)$- function  we conclude from~\eqref{FT} that  in the limit $N\rightarrow\infty$ taken first and a proper choice of $Z$ one has:
\begin{align}
\chi=\lim_{\tau\rightarrow +0}\lim_{N\rightarrow\infty} K(\tau)=\lim_{\Delta\rightarrow \infty}\lim_{N\rightarrow\infty}\chi(\Delta).
\end{align}
 For a finite $N$ the spectral form-factor $K(\tau)$ is a non-trivial function of $\tau$. It increases sharply at $\tau\lesssim (\eta_{0}\,N)^{-1}$, where $\eta_{0}$ is the width of the filter function $\eta(E)$. The sharp increase of $K(\tau)$ at a small $\tau$ seen in Fig.~\ref{fig:fig1}(a) is given by the Fourier transform of the term $Z^{-1}\int dE\,\eta(E)\eta(E-\omega)$ in~\eqref{FT} which is nothing but the broadened $\delta(\tau)$ function discussed above. For the values of $\tau$ in the interval $\delta/E_{{\rm Th}}\lesssim\tau\lesssim 1$ the function $R(\omega)$ is described by the Wigner-Dyson theory: $R(\omega)\propto -1/\omega^{2}$, while $K(\tau)\propto \tau$ grows linearly and approaches the universal limit $K(\tau\gg 1)=1$ due to the $\delta(\omega)$ function in $R(\omega)$. In between, for $(\eta_{0}\,N)^{-1}\ll \tau\ll \delta/E_{{\rm Th}}$, there is a plateau for the critical $K(\tau)$ which height is given by~\eqref{chi-fin}.
 
\subsection*{Derivation of Eq.~(\ref{eq:ans-chi-orth}) and Eq.~(\ref{eq:ans-chi-unit})}
The derivation of~\eqref{eq:ans-chi-unit} for the unitary symmetry class is based on~\eqref{chi-fin} and the global DoS correlation function computed in Ref.~\cite{KrMutta},
 \begin{align}\label{R-F}
R(\omega)=\delta(\omega)-F(\omega)^{2}\;,
 \end{align}
where the limit $N\rightarrow\infty$ is already taken and
\begin{align}\label{F-unit}
F(\omega)=\frac{1}{4b}\frac{\sin(\pi\omega)}{\sinh\left(\frac{\pi \omega}{4b}\right)} \;.
\end{align}
Equation~(\ref{F-unit}) was derived using the non-linear sigma model (NLSM)  perturbative approach of Ref.~\cite{AA} extended to the case of the critical PLRB model in Ref.~\cite{KrMutta}. 

By derivation,~\eqref{F-unit} is valid for $\omega\gg 1$ (in units of $\delta$) and for $b\gg 1$ when the NLSM is justified. The correlation function $R(\omega)$ takes a well-known  form in the limit $b\rightarrow \infty$ when it  exactly coincides with the Wigner-Dyson global DoS correlation function~\cite{Mehta} for all values of $\omega$ down to zero.   Thus the perturbative derivation, which is under control only for $\omega\gg 1$, appears to give an exact result for $R(\omega)$ in the Wigner-Dyson unitary case. This fact is known as a "perturbative exactness".

One could hope that~\eqref{F-unit} has the same property of perturbative exactness and is also valid for all $\omega$, provided that $b\gg 1$. And, indeed, it was shown in Ref.~\cite{KrTsvel} that it gives a correct leading term at $\omega/b\ll 1$ found in Ref.~\cite{Mirlin_Kra}. Thus~\eqref{F-unit} gives a correct leading behavior both at $\omega\gg 1$ and in $\omega\ll b$. Since these two domains overlap at $b\gg \omega\gg 1$,~\eqref{F-unit} should give a correct asymptotic behavior for all $\omega$, provided that $b\gg 1$.

Notice also that the DoS correlation function,~\eqref{R-F}, has a structure of the Wigner-Dyson theory $R(\omega>0)=-F(\omega)^{2}$ for the unitary symmetry class, albeit with the modified kernel $F(\omega)$. As a matter of fact, this kernel corresponds to the one for the non-interacting fermions in 1$d$ at a finite temperature,
\begin{align}
    T=\frac{1}{4b}\;,
\end{align}
while the standard Wigner-Dyson kernel corresponds to zero temperature.

What remains to be done to find $\chi$ in the unitary ensemble is to integrate $R(\omega)$ as in~\eqref{chi-fin}. So we obtain~\eqref{eq:ans-chi-unit}. 

As was explained above, the correctness of the result is guaranteed only for $b\gg 1$. However,~\eqref{eq:ans-chi-unit} describes $\chi$  as a function of $b$ extremely well for all values of parameter $b$ in the unitary PLRB model, see Figs.~\ref{fig:fig3}(b) and~\ref{fig:fig4}(a).

In the standard Wigner-Dyson theory of orthogonal symmetry the relation between $R(\omega)$ at $N\rightarrow\infty$ and $F(\omega)$ at $b\rightarrow\infty$ is more complicated~\cite{Mehta},
\begin{align}\label{R-orth-F}
R(\omega>0)=-F(\omega)^{2}-\frac{dF}{d\omega}\int_{\omega}^{\infty}F(x)dx\,,\;\;\; F(-\omega)=F(\omega).
\end{align}
One may compute $R(\omega)$ using the same kernel,~\eqref{F-unit},  and~\eqref{R-orth-F} and check if it gives a correct leading term in $R(\omega)$ at small $\omega<1$. It was done in Ref.~\cite{KrTsvel} with the affirmative answer. This demonstrates that the status of~\eqref{R-orth-F} with the kernel,~\eqref{F-unit}, is the same as for the unitary ensemble: it is valid for all $\omega$ in the limit $b\gg 1$.

Let us compute $\chi$ from~\eqref{chi-fin} using $R(\omega)$ found as described above. It is convenient first to integrate by parts over $\omega$ and then integrate over $x$ in~\eqref{R-orth-F}. So we obtain:
\begin{align}
2\int_{+0}^{\infty}R(\omega)\,d\omega=2\int_{0}^{\infty}F(x) dx -4\int_{0}^{\infty}F^{2}(x) dx\;.
\end{align}
Then, using
\begin{align}
    \int_{0}^{\infty}F(x) dx=\frac{1}{2}\tanh(2\pi b),\\
     \int_{0}^{\infty}F^{2}(x) dx=-\frac{1}{8\pi b}+\frac{1}{2}\coth(4\pi b),
\end{align}
one obtains~\eqref{eq:ans-chi-orth} from~\eqref{chi-fin}.

By derivation, the status of~\eqref{eq:ans-chi-orth} should be the same as~\eqref{eq:ans-chi-unit}: it should be valid asymptotically at $b\gg 1$. And, indeed, Fig.~\ref{fig:fig4} shows that this is the case. However, for intermediate values of $b$ the agreement with numerical results is not as good as for the unitary ensemble. 

}

\showmatmethods{} 

\acknow{We are grateful to I. M.~Khaymovich and P.~Prelov\v sek for discussions.
S.J.~acknowledges discussions with P.~\L yd\.{z}ba about the numerical implementation of the level number variance, as well as discussions with R.~Pintar and K.~Ogane about the implementation of the Polynomial Filtered Exact Diagonalization method and the cost function minimization approach. 
We acknowledge support from the Slovenian Research and Innovation Agency (ARIS), Research core funding Grants No.~P1-0044, N1-0273 and J1-50005, as well as the Consolidator Grant Boundary-101126364 of the European Research Council (ERC) (S.J, M.H. and L.V.).
M. H. acknowledges support from the Polish National Agency for Academic Exchange (NAWA)’s Ulam Programme (project BNI/ULM/2024/1/00124).
We gratefully acknowledge the High Performance Computing Research Infrastructure Eastern Region (HCP RIVR) consortium~\cite{vega1} and European High Performance Computing Joint Undertaking (EuroHPC JU)~\cite{vega2}  for funding this research by providing computing resources of the HPC system Vega at the Institute of Information sciences~\cite{vega3}.
}

\showacknow{} 


\bibliography{pnas-bib}
\end{document}